\documentclass[11pt,a4paper]{article}
\usepackage{jcappub}
\usepackage{euscript,epsfig,amsmath,amssymb}
\usepackage{amsfonts,latexsym}
\usepackage{color}

\newcommand{\be}[1]{\begin{equation}\label{#1}}
\newcommand{\ee}{\end{equation}}
\newcommand{\ba}[1]{\begin{eqnarray}\label{#1}}
\newcommand{\ea}{\end{eqnarray}}
\newcommand{\rf}[1]{(\ref{#1})}
\newcommand{\nn}{\nonumber}
\renewcommand{\theequation}{\arabic{section}.\arabic{equation}}

\newcommand{\ov}{\overline}

\newcommand{\const}{\mbox{\rm const}\,}

\begin{document}

\title{Scalar and vector perturbations in a universe with discrete and continuous matter sources}

\author{Maxim Eingorn$^{1,2}$,}
\author{Claus Kiefer$^2$}
\author{and Alexander Zhuk$^{3}$}

\affiliation{$^{1}$North Carolina Central University, CREST and NASA Research Centers,\\ Fayetteville st. 1801, Durham, North Carolina 27707, U.S.A.\\}

\affiliation{$^{2}$Institute for Theoretical Physics, University of
  Cologne,\\ Z\"ulpicher Strasse 77, 50937 K\"oln, Germany\\}

\affiliation{$^{3}$Astronomical Observatory, Odessa National University,\\ Dvoryanskaya st. 2, Odessa 65082, Ukraine\\}

\emailAdd{maxim.eingorn@gmail.com}
\emailAdd{kiefer@thp.uni-koeln.de}
\emailAdd{ai.zhuk2@gmail.com}

\abstract{We study a universe filled with dust-like matter in the form of discrete inhomogeneities (e.g., galaxies and their groups and clusters) and two sets of
perfect fluids with linear and nonlinear equations of state, respectively. The background spacetime geometry is defined by the FLRW metric. In the weak
gravitational field limit, we develop the first-order scalar and vector cosmological perturbation theory. Our approach works at all cosmological scales (i.e.
sub-horizon and super-horizon ones) and incorporates linear and nonlinear effects with respect to energy density fluctuations. We demonstrate that the scalar
perturbation (i.e. the gravitational potential) as well as the vector perturbation can be split into individual contributions from each matter source. Each of
these contributions satisfies its own equation.  The velocity-independent parts of the individual gravitational potentials are characterized by a finite
time-dependent Yukawa interaction range being the same for each individual contribution. We also obtain the exact form of the gravitational potential and vector
perturbation related to the discrete matter sources. The self-consistency of our approach is thoroughly checked. The derived equations can form the theoretical
basis for numerical simulations for a wide class of cosmological models.}

\maketitle

\flushbottom

\section{\label{sec:1}Introduction}

The explanation of the accelerated expansion of the late Universe is one of the greatest challenges in modern cosmology. The conventional $\Lambda$CDM model
conforms with the data of the \textsc{Planck} mission
\cite{Planck1,Planck2}. The nature of the cosmological
constant is, however, still unclear. There are, in fact, about 120 orders of
magnitude between its observed value and
the theoretically expected one, which is related to the vacuum
energy density. The
cancellation mechanism between the various energy densities
that would reduce this large theoretical value
of the cosmological constant is still a mystery. In addition, the
$\Lambda$CDM model (as
well as a lot of other dark energy models) faces the coincidence
problem, that is the question, why is the cosmological constant at
present of the same order of magnitude as the
energy density of dark matter? To solve these problems, different
dynamical dark energy models were proposed. For example, dark
energy can be modelled by a
barotropic perfect fluid, that is, a fluid whose pressure  is a
function of its energy density only: $p=f(\varepsilon)$ with a proper
equation of state (EoS). The
linear EoS $p=\omega\varepsilon$ with $\omega=\mathrm{const}$ is the simplest example. It is well known that frustrated networks of topological defects (e.g.
cosmic strings and domain walls with $\omega=-1/3$ and $\omega=-2/3$, respectively) have the form of perfect fluids with constant parameters $\omega$
\cite{ZBG,ShellVil,Avelino,Kumar}. A scalar field can also lead to a constant parameter $\omega$ \cite{WL}. This imposes severe restrictions on the form of the
scalar field potential \cite{ZBG,zhuk1996}. In more general models, the
EoS parameter is not a constant; for example, in the
Chevallier-Polarski-Linder (CPL) model
\cite{ChevPol,Linder} the parameter $\omega$ is a linear function of the scale factor. Moreover, the pressure can be a nonlinear function of the energy density,
as is the case, for example, in the Chaplygin gas model
\cite{Kamenshchik2001cp,Bilic2001cg,Bento2002ps}. Obviously, all these
models are viable only when they are
consistent with the observed large-scale structure of the Universe.

 The theoretical study and numerical simulation of the structure
 formation and evolution are
usually performed with the help of perturbation theory. Therefore, it is essential to construct a theory that allows us to investigate the structure of the
Universe at all scales (i.e. sub-horizon and super-horizon ones), and,
ultimately, to determine the nature of dark energy and dark matter.

Such perturbation theories were constructed in the papers \cite{Eingorn1,Eingorn2}. The former paper is devoted to the concordance $\Lambda$CDM model for which
the dust-like matter is considered in the form of discrete inhomogeneities (e.g., galaxies, their groups and clusters). In the latter article, the matter is taken
in the form of a set of perfect fluids with constant EoS parameters. For both of these models, the first-order perturbation approach is valid for arbitrary scales
and incorporates linear and nonlinear effects (with respect to energy density fluctuations) in the weak gravitational field limit. Consequently, it is of interest
to consider the most general model which contains the two forms of matter, that is, discrete inhomogeneities representing cold dark matter (CDM) and perfect
fluids with constant $\omega$, and which also includes additional perfect fluids with nonlinear EoS. We make such a generalization in the present paper and
construct the self-consistent theory of scalar and vector perturbations for this model. Obviously, the CPL and the Chaplygin gas models are particular cases of
this general model.

The paper is structured as follows. In Section 2, we present the system of linearized Einstein equations for the first-order scalar and vector cosmological
perturbations. In Section 3, we demonstrate that the gravitational potential as well as the vector perturbation can be split into individual contributions from
each matter source, and each of these contributions satisfies its own equation. Here, we also obtain the exact solutions for scalar and vector perturbations
corresponding to the discrete matter component. The derived equations demonstrate that gravitational potentials created by fluctuations of each matter source
undergo the Yukawa-type screening. Then we prove in Section 4 that all individual contributions for scalar and vector perturbations satisfy the rest of the
linearized Einstein equations. The main results are summarized in the concluding Section 5. Appendix A is devoted to the energy-momentum tensors for the
considered matter sources in the first-order (with respect to the scalar and vector perturbations) approximation. In Appendix B, we additionally check the
self-consistency of our approach.

\section{\label{sec:2}The model and basic equations}

\setcounter{equation}{0}

We investigate a universe filled with three different forms of matter. The first form represents discrete gravitating sources (galaxies and their groups with
baryonic and CDM constituents). We consider them at distances much larger than their characteristic sizes. Therefore, they can be treated as point-like masses.
Obviously, this form of matter has a dust-like EoS $p_M=0$ with the average energy density $\overline{\varepsilon}_M=\overline\rho_Mc^2/a^3$ where
$\overline\rho_M=\const$ is the average comoving rest mass density. Second, we consider an arbitrary number of continuous perfect fluids with the linear EoS
$p_I=\omega_I\varepsilon_I,\quad \omega_I=\mathrm{const}\neq0$. This EoS preserves its form for the average values:
$\overline{p}_I=\omega_I\overline{\varepsilon}_I$. Fluctuations of the nonrelativistic matter component ``M'' and ``I''-components are treated in a
non-perturbative way: fulfilment of the inequalities $\delta\varepsilon_M\ll\overline\varepsilon_M$ and $\delta\varepsilon_I\ll\overline\varepsilon_I$  are not
demanded; for instance, the intragalactic medium and dark matter halos are characterized by values of $\varepsilon_M$ much higher than $\overline\varepsilon_M$.
The third type of perfect fluids is characterized by a nonlinear EoS $p_J=f_J(\varepsilon_J)$, where $f_J$ are some nonlinear functions. Clearly, the CPL and the
Chaplygin gas models are  particular examples of such a perfect fluid. We can expand these EoS near the average values $\overline\varepsilon_J$:
$p_J=f_J(\overline\varepsilon_J)+(\partial f_J/\partial\overline\varepsilon_J)\delta\varepsilon_J +(1/2)(\partial^2
f_J/\partial\overline\varepsilon_J^2)\delta\varepsilon_J^2 +\cdots$. Since $\overline{\delta\varepsilon_J}=0$, for the averaged EoS we get $\overline
p_J=f_J(\overline\varepsilon_J)+(1/2)(\partial^2 f_J/\partial\overline\varepsilon_J^2){\overline{\delta\varepsilon_J^2}} +\cdots$. Hence, if we demand
$\delta\varepsilon_J/\overline\varepsilon_J=o(1)$, we can drop small correction terms of the order $O({\overline{\delta\varepsilon_J^2}}) $ and write
(approximately) the EoS $\overline p_J\approx f_J(\overline\varepsilon_J)$ for the average values.

For the considered model, the background Friedmann equations (for flat spatial sections) read
\be{2.1} \frac{3{\mathcal H}^2}{a^2}=\frac{3H^2}{c^2}=\kappa\left(\overline{\varepsilon}_M+\sum\limits_I\overline{\varepsilon}_I+
\sum\limits_J\overline{\varepsilon}_J\right)\, ,\ee
\be{2.2} \frac{2{\mathcal H}'+{\mathcal H}^2}{a^2}= \frac{1}{c^2}\left(3H^2+2\dot H\right)=\frac{H^2}{c^2}\left(1-2q\right)=
-\kappa\left(\sum\limits_I\overline{p}_I+\sum\limits_J\overline{p}_J\right)\, , \ee
with the EoS
\be{2.3}
\ov p_M=0, \qquad \ov p_I=\omega_I\ov \varepsilon_I,\, \, \omega_I=\mathrm{const}\neq0,\qquad
\overline p_J=f(\overline\varepsilon_J),\, \, \delta\varepsilon_J/\overline\varepsilon_J=o(1)\, .
\ee
In Eqs. \rf{2.1} and \rf{2.2}, the Hubble parameters are $\mathcal H\equiv a'/a\equiv (da/d\eta)/a$ and $H\equiv \dot a/a\equiv (da/d t)/a$. Hereafter, the prime
and overdot denote the derivatives with respect to the conformal ($\eta$) and synchronous ($t$) times, respectively. They are connected by $c dt = a d\eta$, where
$c$ is the speed of light and $a$ is the scale factor. The constant $\kappa\equiv 8\pi G_N/c^4$ is introduced as well ($G_N$ is the Newtonian gravitational
constant). Eqs. \rf{2.1} and \rf{2.2} lead to a useful auxiliary relation:
\be{2.4} \frac{3}{2}\kappa\left[\overline{\varepsilon}_M+\sum\limits_I(\overline{\varepsilon}_I+\ov p_I)+\sum\limits_J(\overline{\varepsilon}_J+\ov p_J)\right]=
\frac{3}{a^2}\left(\mathcal{H}^2-\mathcal{H}'\right)=\frac{3}{c^2}H^2(1+q)\equiv \frac{1}{\lambda^2}\, , \ee
where $q\equiv-(\ddot a/a)/H^2$ is the deceleration parameter, and we introduced a new variable $\lambda$ which has dimension of length and depends on time. This
variable turns out to be of great importance later.

We now turn to the perturbation equations. The perturbed metric in the first-order approximation is taken in the form \cite{Bardeen,Mukhanov,Durrer,Rubakov}
\be{2.5} ds^2\equiv g_{ik}dx^idx^k\approx a^2\left[\left(1 + 2\Phi\right)d\eta^2 + 2B_{\alpha}dx^{\alpha}d\eta -\left(1-
2\Phi\right)\delta_{\alpha\beta}dx^{\alpha}dx^{\beta}\right]\, . \ee
In the first-order approximation with respect to $\Phi$ and $B_{\alpha}$, the square root of the determinant for this metric is
\be{2.6}
\sqrt{-g}\approx a^4(1-2\Phi)\, .
\ee
For the vector perturbations $B_{\alpha}$, we choose the transverse gauge condition \cite{Durrer2,Green}:
\be{2.7}
\nabla {\bf B}\equiv \delta^{\alpha\beta}\frac{\partial B_{\alpha}}{\partial x^{\beta}}=0\, .
\ee
In the case of the metric \rf{2.5}, the linearized Einstein equations are reduced to the following system of four equations:
\be{2.8} \triangle\Phi-3{\mathcal H}(\Phi'+{\mathcal H}\Phi) =\frac{1}{2}\kappa a^2 \delta T_0^0=\frac{1}{2}\kappa
a^2\left(\frac{c^2}{a^3}\delta\rho_M+\frac{3\overline\rho_M c^2}{a^3}\Phi +\sum\limits_I\delta{\varepsilon}_I+\sum\limits_J\delta{\varepsilon}_J\right)\, , \ee
\ba{2.9} &{}&\frac{1}{4}\triangle{B}_{\alpha}+\frac{\partial}{\partial x^{\alpha}}(\Phi'+{\mathcal H}\Phi)=\frac{1}{2}\kappa a^2\delta
T_{\alpha}^0=\frac{1}{2}\kappa
a^2\left(-\frac{c^2}{a^3}\sum\limits_n\rho_n\tilde v^{\alpha}_n+\frac{\overline\rho_M c^2}{a^3}B_{\alpha}\right.\nn\\
&-&\left.\sum\limits_I(\varepsilon_I+p_I)\tilde
v_I^{\alpha} +
\sum\limits_I(\overline\varepsilon_I+\overline p_I)B_{\alpha}-\sum\limits_J(\overline\varepsilon_J+\overline p_J)\tilde v_J^{\alpha}+
\sum\limits_J(\overline\varepsilon_J+\overline p_J)B_{\alpha}\right)\, ,
\ea
\be{2.10}
\Phi''+3{\mathcal H}\Phi'+\left(2{\mathcal H}'+{\mathcal H}^2\right)\Phi=\frac{1}{2}\kappa
a^2\left(\sum\limits_I\delta{p}_I+\sum\limits_J\delta{p}_J\right)\, ,
\ee
\be{2.11}
\left(\frac{\partial B_{\alpha}}{\partial x^{\beta}}+\frac{\partial B_{\beta}}{\partial x^{\alpha}}\right)'+2{\mathcal H}\left(\frac{\partial
B_{\alpha}}{\partial x^{\beta}}+\frac{\partial B_{\beta}}{\partial x^{\alpha}}\right)=0\, .
\ee
Here, $\triangle$ is the Laplace operator in flat space. On the right-hand side of these equations, we used the linearized expressions for the energy-momentum
tensors given by Eqs. \rf{a.7}-\rf{a.10} and \rf{a.15}-\rf{a.17}. As defined in Appendix A, $\rho_n$ is the rest mass density of a gravitating source with mass
$m_n$, and $\tilde v^{\alpha}_n$ is the peculiar velocity of this discrete source. The mixed $\alpha\beta$ component of the linearized Einstein equation was split
into scalar \rf{2.10} and vector \rf{2.11} parts. Additionally, for the linearized components of the energy-momentum tensor we took into account the following.
Eqs. \rf{2.8}-\rf{2.10} indicate that $\delta\rho_M, \delta \varepsilon_{J,I}$ and $\delta p_{J,I}$ are already sources for the metric corrections $\Phi$ and
$B_{\alpha}$. Hence, the products of $\delta\rho_M, \delta \varepsilon_{J,I}$ and $\delta p_{J,I}$ with $\Phi$ and $B_{\alpha}$ result in corrections of second
order. Therefore, in all products of the form $\rho_M,\varepsilon_{J,I},p_{J,I} \times \Phi, B_{\alpha}$ we replace $\rho_M, \varepsilon_{J,I}$ and $p_{J,I}$ by
their average values $\ov\rho_M, \ov\varepsilon_{J,I}$ and $\ov p_{J,I}$. For example, $\rho_M\Phi \rightarrow \ov\rho_M\Phi,\, \, p_{I}B_{\alpha} \rightarrow \ov
p_{I}B_{\alpha}$, etc. On the other hand, the peculiar velocities appear as sources of the metric corrections only in the combinations $\rho_n\tilde v^{\alpha}_n$
and $(\varepsilon_{J,I}+p_{J,I})\tilde v^{\alpha}_{J,I}$. These combinations result in corrections of first order and we should thus preserve them. However,
fluctuations of the energy density and pressure of the ``J''-components (in contrast to the ``M''- and ``I''-components) are small: $\delta\varepsilon_J,\delta
p_J \ll \bar\varepsilon_J, \bar p_J$. Therefore, we can replace $(\varepsilon_J+ p_J)\tilde v_J^{\alpha}$ by $(\overline\varepsilon_J+\overline p_J)\tilde
v_J^{\alpha}$. In the following equations, we will use the reasoning described in this paragraph.

Taking into account these comments, the conservation equation \rf{a.20} for the ``I''-components reads
\be{2.12}
\varepsilon'_I+3\mathcal{H}(1+\omega_I)\varepsilon_I-3(1+\omega_I)\overline\varepsilon_I\Phi'+(1+\omega_I)\nabla\left(\varepsilon_I\tilde{\bf
v}_I\right)=0\, ,
\ee
where we used the replacement $\nabla[p_I \mathbf{B}]= \nabla[(\ov p_I+\delta p_I) \mathbf{B}] \rightarrow\bar p_I\nabla \mathbf{B}=0$. Obviously, the
``M''-component also satisfies this equation with $\omega_I=0$. We are looking for a solution of this equation in the form
\be{2.13}
\varepsilon_I=\frac{A_I}{a^{3(1+\omega_I)}}+3(1+\omega_I)\overline\varepsilon_I\Phi=\frac{\overline A_I}{a^{3(1+\omega_I)}}+\frac{\delta
A_I}{a^{3(1+\omega_I)}}+\frac{3(1+\omega_I)\overline A_I}{a^{3(1+\omega_I)}}\Phi\, ,
\ee
where $\ov A_I =$ const. Therefore,
\be{2.14}
\overline\varepsilon_I=\frac{\overline A_I}{a^{3(1+\omega_I)}}
\ee
and
\be{2.15}
\delta\varepsilon_I=\frac{\delta A_I}{a^{3(1+\omega_I)}}+\frac{3(1+\omega_I)\overline
A_I}{a^{3(1+\omega_I)}}\Phi\, , \quad \delta A_I\equiv A_I-\overline A_I\, .
\ee
The function $A_I$ satisfies the equation
\be{2.16}
A'_I+(1+\omega_I)\nabla\left(A_I\tilde{\bf v}_I\right)=0\, .
\ee
In the case of the ``M''-component, we should perform the obvious substitutions, e.g., $\omega_I=0,\, A_I \rightarrow \rho_M c^2=\sum_n\rho_n c^2,\, \ov A_I
\rightarrow \ov \rho_M c^2, \, \delta A_I \rightarrow \delta\rho_M
c^2$. For instance, Eq. \rf{2.16} is then replaced by
\be{2.17}
\rho'_n+\nabla\left(\rho_n\tilde{\bf v}_n\right)=0\, .
\ee
Eqs. \rf{2.16} and \rf{2.17} demonstrate that the quantities $A_I$ and
$\rho_n$ do not depend on time in the case when the peculiar
velocities are absent, $\tilde{\bf
v}_I,\tilde{\bf v}_n=0$, that is, they become pure comoving.
With respect to the ``J''-components, Eq. \rf{a.20} reads
\be{2.18}
\varepsilon'_J+3\mathcal{H}(\varepsilon_J+p_J)-3(\overline\varepsilon_J+\overline p_J)\Phi'+(\overline\varepsilon_J+\overline p_J)(\nabla\tilde{\bf
v}_J)=0\, .
\ee
We can split $\varepsilon_J$ into background and perturbation parts: $\varepsilon_J=\ov \varepsilon_J+\delta \varepsilon_J$. These parts satisfy the following
equations:
\be{2.19}
\overline\varepsilon'_J+3\mathcal{H}(\overline\varepsilon_J+\overline p_J)=0
\ee
and
\be{2.20}
\delta\varepsilon'_J+3\mathcal{H}(\delta\varepsilon_J+\delta
p_J)-3(\overline\varepsilon_J+\overline p_J)\Phi'+(\overline\varepsilon_J+\overline p_J)(\nabla\tilde{\bf v}_J)=0\, .
\ee
Assuming that
\be{2.21}
\delta\varepsilon_J=\overline\varepsilon_J\delta_J+3(\overline\varepsilon_J+\overline p_J)\Phi\, ,
\ee
we find that a new function $\delta_J$ defined by this formula satisfies the equation
\be{2.22}
\overline\varepsilon_J\delta'_J+3\mathcal{H}\left(\frac{d\overline p_J}{d\overline\varepsilon_J}\overline\varepsilon_J-\overline
p_J\right)\delta_J+(\overline\varepsilon_J+\overline p_J)(\nabla\tilde{\bf v}_J)=0\, .
\ee
Substituting \rf{2.13} into \rf{2.9}, we obtain
\ba{2.23} &{}&\frac{1}{4}\triangle{B}_{\alpha}+\frac{\partial}{\partial x^{\alpha}}(\Phi'+{\mathcal H}\Phi)=\frac{1}{2}\kappa
a^2\left(-\frac{c^2}{a^3}\sum\limits_n\rho_n\tilde v^{\alpha}_n+\frac{\overline\rho_M c^2}{a^3}B_{\alpha}-
\sum\limits_I\frac{1+\omega_I}{a^{3(1+\omega_I)}}A_I\tilde
v_I^{\alpha}\right.\nn\\
&+&\left.\sum\limits_I(\overline\varepsilon_I+\overline p_I)B_{\alpha}-\sum\limits_J(\overline\varepsilon_J+\overline p_J)\tilde v_J^{\alpha}+
\sum\limits_J(\overline\varepsilon_J+\overline p_J)B_{\alpha}\right)\, ,
\ea
where we dropped the term $\Phi\tilde v_I^{\alpha}$. Now,
 we present all terms (on the right-hand side of this equation) containing the vectors of the peculiar velocities
as the sum of the longitudinal and transverse parts, that is, as the sum
of the gradient and the curl:
\ba{2.24}
\sum_n\rho_n\tilde {\bf v}_n &=& \nabla \Xi+\left(\sum_n\rho_n\tilde {\bf v}_n-\nabla\Xi\right)\, ,
\quad \nabla \left(\sum_n\rho_n\tilde {\bf v}_n\right)=\triangle\Xi\, ,\\
\label{2.25}A_I\tilde {\bf v}_I&=&\nabla\xi_I+\left(A_I\tilde {\bf v}_I-\nabla\xi_I\right)\, ,
\quad \qquad \nabla(A_I\tilde {\bf v}_I)=\triangle\xi_I\, ,\\
\label{2.26} \tilde {\bf v}_J&=&\nabla \nu_J+\left(\tilde {\bf v}_J-\nabla \nu_J\right)\, , \quad \quad \quad \ \ \, \nabla\tilde {\bf v}_J=\triangle \nu_J\, .
\ea
Here, the function $\nu_J$ is the velocity potential of the
``J''-component. The functions $\Xi$ and $\xi_I$ can also be treated
as effective velocity potentials
of the ``M''- and ``I''-components, respectively. The function $\Xi$
reads \cite{Eingorn1}
\be{2.27}
\Xi=\frac{1}{4\pi}\sum\limits_nm_n\frac{({\bf r}-{\bf r}_n)\tilde {\bf v}_n}{|{\bf r}-{\bf r}_n|^3}\, .\ee
Taking into account \rf{2.24}-\rf{2.26}, Eq. \rf{2.23} can be split into scalar and vector parts (longitudinal and transverse parts):
\be{2.28}
\Phi'+{\mathcal H}\Phi=-\frac{\kappa c^2}{2a}\Xi-\frac{\kappa}{2}\sum\limits_I\frac{1+\omega_I}{a^{1+3\omega_I}}\xi_I-\frac{\kappa
a^2}{2}\sum\limits_J(\overline\varepsilon_J+\overline p_J)\nu_J
\ee
and
\ba{2.29}
&{}&\frac{1}{4}\triangle{\bf B}-\frac{\kappa\overline\rho c^2}{2a}{\bf B}=-\frac{\kappa c^2}{2a}\left(\sum\limits_n\rho_n\tilde{\bf v}_n-\nabla\Xi\right)
-\frac{\kappa}{2}\sum\limits_I\frac{1+\omega_I}{a^{1+3\omega_I}}\left(A_I\tilde {\bf v}_I-\nabla\xi_I\right)\nn\\
&-&\frac{\kappa a^2}{2}\sum\limits_J(\overline\varepsilon_J+\overline p_J)\left(\tilde {\bf v}_J-\nabla \nu_J\right)+\frac{\kappa
a^2}{2}\sum\limits_I(\overline\varepsilon_I+\overline p_I){\bf B}+\frac{\kappa a^2}{2}\sum\limits_J(\overline\varepsilon_J+\overline p_J){\bf B}\, .
\ea

\section{\label{sec:3}Equations for
gravitational potential $\Phi$ and
  vector perturbation ${\bf B}$}

\setcounter{equation}{0}

\subsection{Equations for gravitational potential $\Phi$}

The aim of this subsection is to derive separate equations for
  the gravitational potentials from three considered forms of matter and to
  derive an exact solution for the case of discrete sources.
Substituting \rf{2.28} into \rf{2.8}, we get
\ba{3.1} \triangle\Phi &=& \frac{1}{2}\kappa a^2\left(\frac{c^2}{a^3}\delta\rho_M+\frac{3\overline\rho_M c^2}{a^3}\Phi
+\sum\limits_I\delta{\varepsilon}_I+\sum\limits_J\delta{\varepsilon}_J\right)\nn\\
&-&\frac{3\mathcal{H}\kappa c^2}{2a}\Xi-\frac{3\mathcal{H}\kappa}{2}\sum\limits_I\frac{1+\omega_I}{a^{1+3\omega_I}}\xi_I-\frac{3\mathcal{H}\kappa
a^2}{2}\sum\limits_J(\overline\varepsilon_J+\overline p_J)\nu_J\, .
\ea
After taking into account Eqs. \rf{2.15} and \rf{2.21}, we get
\ba{3.2}
&{}&\triangle\Phi-\frac{a^2}{\lambda^2}\Phi =\frac{\kappa c^2}{2a}\delta\rho_M+\frac{\kappa a^2}{2}\sum\limits_I\frac{\delta
A_I}{a^{3(1+\omega_I)}}+\frac{\kappa a^2}{2}\sum\limits_J\overline\varepsilon_J\delta_J\nn\\
&-&\frac{3\kappa c^2{\mathcal H}}{2a}\Xi-\frac{3\mathcal{H}\kappa}{2}\sum\limits_I\frac{1+\omega_I}{a^{1+3\omega_I}}\xi_I-\frac{3\mathcal{H}\kappa
a^2}{2}\sum\limits_J(\overline\varepsilon_J+\overline p_J)\nu_J\, ,
\ea
where the variable parameter $\lambda$ defined in \rf{2.4} now reads
\be{3.3}
\lambda=\left[\frac{3\kappa\overline\rho_M c^2}{2a^3}+\frac{3\kappa}{2}\sum\limits_I\frac{(1+\omega_I)\overline
A_I}{a^{3(1+\omega_I)}}+\frac{3\kappa}{2}\sum\limits_J(\overline\varepsilon_J+\overline p_J)\right]^{-1/2}\, .
\ee
If we want to consider only scalar perturbations, then, in all equations, we must ignore the vector perturbation $\mathbf{B}$ and the curls defined in Eqs.
\rf{2.24}-\rf{2.26}. In this case, the left-hand side of these equations are defined only by their gradient parts, that is, by the velocity potentials for
\rf{2.26} and by the effective velocity potentials for \rf{2.24} and \rf{2.25}. It is worth noting that Eq. \rf{3.2} agrees with Eq. (26) in \cite{Eingorn2},
which represents the particular case of Eq.~\rf{3.2}.


It makes sense to split the total gravitational potential $\Phi$ into individual contributions from each matter source:
\be{3.4}
\Phi=\Phi_M+\Phi_{\Sigma I}+\Phi_{\Sigma J},\quad \Phi_{\Sigma I}\equiv\sum\limits_I\Phi_I,\quad \Phi_{\Sigma J}\equiv\sum\limits_J\Phi_J\, ,\ee
where these individual gravitational potentials satisfy the following equations:
\ba{3.5}
\triangle\Phi_M-\frac{a^2}{\lambda^2}\Phi_M &=&\frac{\kappa c^2}{2a}\delta\rho_M-\frac{3\kappa c^2{\mathcal H}}{2a}\Xi\, ,\\
\label{3.6} \triangle\Phi_{I}-\frac{a^2}{\lambda^2}\Phi_{I} &=&\frac{\kappa}{2}\frac{\delta
A_I}{a^{1+3\omega_I}}-\frac{3\mathcal{H}\kappa}{2}\frac{1+\omega_I}{a^{1+3\omega_I}}\xi_I\, ,\\
\label{3.7} \triangle\Phi_{J}-\frac{a^2}{\lambda^2}\Phi_{J}& =&\frac{\kappa a^2}{2}\overline\varepsilon_J\delta_J-\frac{3\mathcal{H}\kappa
a^2}{2}(\overline\varepsilon_J+\overline p_J)\nu_J\, .
\ea
We note that Eq. \rf{3.5} coincides with (2.27)
in \cite{Eingorn1} (with the corresponding redefinition of $\lambda$), while Eq. \rf{3.6} agrees with (29) in \cite{Eingorn2}.


Eq. \rf{3.5} for the gravitational potentials of the discrete sources (gravitating masses) can be solved exactly because $\delta\rho_M$ and $\Xi$ have a rather
simple form in momentum space where Eq. \rf{3.5} reads\footnote{Hereafter, the hats denote the Fourier transforms.}
\be{3.8}
 -k^2\hat\Phi_M-\frac{a^2}{\lambda^2}\hat\Phi_M =\frac{\kappa c^2}{2a}\hat{\delta\rho}_M-\frac{3\kappa c^2{\mathcal H}}{2a}\hat\Xi\, ,
\ee
where (see \cite{Eingorn1})
\be{3.9}
\hat{\delta\rho}_M=\sum\limits_nm_n\exp(-i{\bf k}{\bf r}_n)-\overline\rho_M(2\pi)^3\delta({\bf k}),\quad \hat\Xi=-\frac{i}{k^2}\sum\limits_nm_n({\bf
k}\tilde{\bf v}_n)\exp(-i{\bf k}{\bf r}_n)\, .
\ee
Therefore,
\be{3.10} \hat\Phi_M =-\frac{\kappa c^2}{2a}\left(k^2+\frac{a^2}{\lambda^2}\right)^{-1}\left[\sum\limits_nm_n\exp(-i{\bf k}{\bf r}_n)\left(1+3i{\mathcal
H}\frac{({\bf k}\tilde{\bf v}_n)}{k^2}\right)-\overline\rho_M(2\pi)^3\delta({\bf k})\right]\, ,\ee
and the inverse Fourier transformation gives
\ba{3.11}
\Phi_M&=&\frac{\kappa\overline\rho_M c^2\lambda^2}{2a^3}-\frac{\kappa c^2}{8\pi a}\sum\limits_n\frac{m_n}{|{\bf r}-{\bf r}_n|}\exp(-q_n)\nn\\
&+&\frac{3\kappa c^2}{8\pi a}{\mathcal H}\sum\limits_n \frac{m_n[\tilde{\bf v}_n({\bf r}-{\bf r}_n)]}{|{\bf r}-{\bf r}_n|}\cdot\frac{1-(1+q_n)\exp(-q_n)}{q_n^2}\, ,
\ea
where
\be{3.12}
{\bf q}_n(\eta,{\bf r}) \equiv \frac{a}{\lambda}({\bf r}-{\bf r}_n)=\frac{1}{\lambda}({\bf R}-{\bf R}_n),\quad q_n\equiv|{\bf q}_n|\, .
\ee
With the proper redefinition of the parameter $\lambda$, Eq. \rf{3.11}
coincides with the expression (2.40) in \cite{Eingorn1}. Following the
line of the paper
\cite{Eingorn1}, we can show that the average value of $\Phi_M$ when
averaging over the whole Universe is equal to zero:
$\overline\Phi_M=0$, as it should be.

The physical meaning of the parameter $\lambda$ follows from Eqs. \rf{3.2}, \rf{3.5}-\rf{3.7}. This parameter defines the range of the Yukawa interaction. We can
see it explicitly in Eq. \rf{3.11}.

It is worth noting that the Yukawa-type screening of the gravitational potentials of inhomogeneities takes place also within the mechanical approach proposed in
\cite{EZcosm1,EZcosm2,EKZ2}. In this approach, the velocities of the inhomogeneities and fluctuations of other perfect fluids are neglected. Then, these perfect
fluids find themselves in a very specific coupled state: their energy density and pressure fluctuations are proportional to the gravitational potential. In other
words, they are concentrated around the inhomogeneities (e.g., galaxies). For such perfect fluids, the fluctuations $\delta A_I,\delta_J=0$ in Eqs. \rf{2.15} and
\rf{2.21}, respectively, and the equation for the gravitational potential has the form of Eq. \rf{3.2} where we should additionally neglect the velocity
potentials and redefine correspondingly the expression \rf{3.3} for $\lambda$ (see, e.g., \cite{BUZ1,coupled}).

We note that coupled states of perfect fluids were investigated in the case of a perfect fluid with a constant equation of state parameter \cite{BUZ1} and for the
following cosmological scenarios and constituents of the Universe: quark-gluon nuggets \cite{Laslo2}, the CPL model \cite{CPL}, Chaplygin gas \cite{Chaplygin},
nonlinear $f(R)$ gravity \cite{Novak}, as well as the models with a scalar field \cite{scalfield,quintessence} and dark sector interactions
\cite{Kiefer,MariamJoao}.

The appearance of the Yukawa-type screening of the gravitational potentials for each matter source is a very important feature. It is clear from the physical
point of view that at scales exceeding the range of the Yukawa interaction (the time-dependent parameter $\lambda$ in our case), the inhomogeneities are very
weakly gravitationally bound. This means that the largest structures in the Universe should be of the order of $\lambda$ \cite{Eingorn1,Eingorn2}.  In the case of
the $\Lambda$CDM model, the current value of $\lambda$ is estimated as $\lambda\approx 3.7$ Gpc \cite{Eingorn1}. Remarkably, this value is of the order of the
largest observed structures in the present Universe \cite{supstr1,supstr2,supstr3}.

\subsection{Equations for vector perturbation ${\bf B}$}

It can be easily seen that Eq. \rf{2.29} may be rewritten as
\ba{3.13}
&{}&\frac{1}{4}\triangle{\bf B}-\frac{a^2}{3\lambda^2}{\bf B}=-\frac{\kappa c^2}{2a}\left(\sum\limits_n\rho_n\tilde{\bf v}_n-\nabla\Xi\right)
\nn\\&-&\frac{\kappa}{2}\sum\limits_I\frac{1+\omega_I}{a^{1+3\omega_I}}\left(A_I\tilde {\bf v}_I-\nabla\xi_I\right)-\frac{\kappa
a^2}{2}\sum\limits_J(\overline\varepsilon_J+\overline p_J)\left(\tilde {\bf v}_J-\nabla \nu_J\right)\, .
\ea
By analogy with the gravitational potential $\Phi$, we can split the total vector perturbation ${\bf B}$ into individual contributions from each matter source:
\be{3.14}
{\bf B}={\bf B}_M+{\bf B}_{\Sigma I}+{\bf B}_{\Sigma J},\quad {\bf B}_{\Sigma I}\equiv\sum\limits_I{\bf B}_I,\quad {\bf B}_{\Sigma
J}\equiv\sum\limits_J{\bf B}_J\, ,
\ee
where each individual contribution satisfies the following equations:
\ba{3.15}
\frac{1}{4}\triangle{\bf B}_M-\frac{a^2}{3\lambda^2}{\bf B}_M&=&-\frac{\kappa c^2}{2a}\left(\sum\limits_n\rho_n\tilde{\bf v}_n-\nabla\Xi\right)\, ,\\
\label{3.16} \frac{1}{4}\triangle{\bf B}_{I}-\frac{a^2}{3\lambda^2}{\bf B}_{I}&=&-\frac{\kappa}{2}\frac{1+\omega_I}{a^{1+3\omega_I}}\left(A_I\tilde {\bf
v}_I-\nabla\xi_I\right)\, ,\\
\label{3.17} \frac{1}{4}\triangle{\bf B}_{J}-\frac{a^2}{3\lambda^2}{\bf B}_{J}&=&-\frac{\kappa a^2}{2} (\overline\varepsilon_J+\overline p_J)\left(\tilde {\bf
v}_J-\nabla \nu_J\right)\, . \ea
The right-hand sides of these equations are the curls. Therefore, the gauge condition \rf{2.7} is satisfied: ${\bf k}\hat{\bf B}_M=0$, ${\bf k}\hat{\bf B}_I=0$
and ${\bf k}\hat{\bf B}_J=0$. We can find the explicit analytic expression for ${\bf B}_M$. To do it, we write Eq. \rf{3.15} in momentum space:
\be{3.18}
-\frac{k^2}{4}\hat{\bf B}_M-\frac{a^2}{3\lambda^2}\hat{\bf B}_M=-\frac{\kappa c^2}{2a}\left(\sum\limits_n\hat\rho_n\tilde{\bf v}_n-i{\bf
k}\hat\Xi\right)\, ,
\ee
where the Fourier transform $\hat\rho_n$ is (see \cite{Eingorn1})
\be{3.19}
\hat\rho_n=m_n\exp(-i{\bf k}{\bf r}_n)\, .
\ee
Therefore, taking into account the second expression in Eq.~\rf{3.9}, we get
\be{3.20}
\hat{\bf B}_M=\frac{2\kappa c^2}{a}\left(k^2+\frac{4a^2}{3\lambda^2}\right)^{-1}\sum\limits_nm_n\exp(-i{\bf k}{\bf r}_n)\left(\tilde{\bf v}_n-\frac{({\bf
k}\tilde{\bf v}_n)}{k^2}{\bf k}\right)\, .
\ee
The inverse Fourier transformation gives
\ba{3.21}
{\bf B}_M&=& \frac{\kappa c^2}{8\pi a}\sum\limits_{n}\left[\frac{m_n\tilde{\bf v}_n}{|{\bf r}-{\bf r}_n|}\cdot
\frac{(3+2\sqrt{3}q_n+4q_n^2)\exp(-2q_n/\sqrt{3})-3}{q_n^2}\right.\nn\\
&+&\left.\frac{m_n[\tilde{\bf v}_n({\bf r}-{\bf r}_n)]}{|{\bf r}-{\bf r}_n|^3}({\bf r}-{\bf
r}_n)\cdot\frac{9-(9+6\sqrt{3}q_n+4q_n^2)\exp(-2q_n/\sqrt{3})}{q_n^2}\right]\, .
\ea
This expression exactly coincides with the formula (2.36) in \cite{Eingorn1}.

\section{\label{sec:4}Two remaining perturbed Einstein equations}

\setcounter{equation}{0}

Now we will demonstrate that two remaining perturbed Einstein equations \rf{2.10} and \rf{2.11} are fulfilled for the metric corrections $\Phi$ and ${\mathbf{B}}$
derived in the previous section. 

\subsection{Check of Equation \rf{2.11} for ${\bf B}$}

To prove Eq. \rf{2.11}, it is sufficient to show that the vector perturbation $\mathbf{B}$ satisfies the condition
\be{4.1}
\mathbf{B}'+2\mathcal{H}\mathbf{B}=0\, .
\ee
First, let us consider the ``M''-component \rf{3.20}. Taking into account that (see Eqs. (2.33)-(2.35) in \cite{Eingorn1})
\ba{4.2} \sum\limits_n\hat\rho_n(a\tilde{\bf v}_n)'=\sum\limits_nm_n\exp(-i{\bf k}{\bf r}_n)(a\tilde{\bf v}_n)'=-a\overline\rho_M\cdot i{\bf
k}\hat\Phi+\overline\rho_M(a\hat{\bf B})'\, ,\ea
we obtain
\ba{4.3} \hat{\bf B}'_M &=&\left[2\kappa c^2\left(a^2k^2+\frac{4a^4}{3\lambda^2}\right)^{-1}\right]' \sum\limits_nm_n\exp(-i{\bf k}{\bf r}_n)\left(a\tilde{\bf
v}_n-\frac{({\bf
k}a\tilde{\bf v}_n)}{k^2}{\bf k}\right)\nn\\
&+&\frac{2\kappa c^2}{a^2}\left(k^2+\frac{4a^2}{3\lambda^2}\right)^{-1}\left[-a\overline\rho_M\cdot i{\bf k}\hat\Phi+\overline\rho_M(a\hat{\bf B})'+\frac{({\bf
k}\cdot a\overline\rho_M\cdot i{\bf k}\hat\Phi)}{k^2}{\bf k}\right]\\
&=&-\left(2k^2+\frac{4}{3a}\frac{d}{da}\left[\frac{a^4}{\lambda^2}\right]\right)\left(k^2+\frac{4a^2}{3\lambda^2}\right)^{-1}\mathcal{H}\hat{\bf B}_M
+\frac{2\kappa \overline\rho_M c^2}{a}\left(k^2+\frac{4a^2}{3\lambda^2}\right)^{-1}(\mathcal{H}\hat{\bf B}+\hat{\bf B}')\nn\, ,
\ea
where we drop the quadratic terms $\mathbf{r}_n'\tilde{\mathbf{v}}_n=\tilde{v}_n^2$. For the ``I''-components in momentum space, we get from Eq. \rf{3.16}
\be{4.4}
\hat{\bf B}_{\Sigma I}=\frac{2\kappa}{a^2}\left(k^2+\frac{4a^2}{3\lambda^2}\right)^{-1}\sum\limits_I(1+\omega_I)a^{1-3\omega_I}\boxed{\left(A_I\tilde {\bf
v}_I-\nabla\xi_I\right)} \, ,
\ee
where a box here and below denotes the Fourier transform. Further, we consider the conservation equation \rf{a.23} which for the ``I''-components reads:
\ba{4.5}
&&\left[\frac{(1+\omega_I)\overline A_I}{a^{3(1+\omega_I)}}{\bf B}\right]'-\left[\frac{(1+\omega_I)}{a^{3(1+\omega_I)}}A_I\tilde{\bf
v}_I\right]'+4\mathcal{H}\frac{(1+\omega_I)\overline A_I}{a^{3(1+\omega_I)}}{\bf B}-4\mathcal{H}\frac{(1+\omega_I) }{a^{3(1+\omega_I)}}A_I\tilde{\bf v}_I\nn\\
&-&\omega_I\nabla \left(\frac{\delta A_I}{a^{3(1+\omega_I)}}+\frac{3(1+\omega_I)\overline A_I}{a^{3(1+\omega_I)}}\Phi\right)-\frac{(1+\omega_I)\overline
A_I}{a^{3(1+\omega_I)}}\nabla\Phi=0\, .\ea
We can use now the formula \rf{2.25} to split this equation into the gradient (longitudinal) and curl (transverse) parts:
\be{4.6}
-\frac{(1+\omega_I)}{a^{3(1+\omega_I)}}\xi_I'-\mathcal{H}(1-3\omega_I)\frac{(1+\omega_I) }{a^{3(1+\omega_I)}}\xi_I -\omega_I\frac{\delta
A_I}{a^{3(1+\omega_I)}}-(1+3\omega_I)\frac{(1+\omega_I)\overline A_I}{a^{3(1+\omega_I)}}\Phi=0
\ee
and
\ba{4.7}
&&\frac{(1+\omega_I)\overline A_I}{a^{3(1+\omega_I)}}{\bf B}'+\mathcal{H}(1-3\omega_I)\frac{(1+\omega_I)\overline A_I}{a^{3(1+\omega_I)}}{\bf
B}\nn\\
&-&\frac{(1+\omega_I)}{a^{3(1+\omega_I)}}\left(A_I\tilde {\bf v}_I-\nabla\xi_I\right)'-\mathcal{H}(1-3\omega_I)\frac{(1+\omega_I)
}{a^{3(1+\omega_I)}}\left(A_I\tilde {\bf v}_I-\nabla\xi_I\right)=0\, .
\ea
Therefore,  from Eqs. \rf{4.4} and \rf{4.7} we get
\ba{4.8} \hat{\bf B}'_{\Sigma I} &=& \left[-2\kappa\left(2a^2k^2+\frac{4}{3}a\frac{d}{da}\left[\frac{a^4}{\lambda^2}\right]\right)
\left(a^2k^2+\frac{4a^4}{3\lambda^2}\right)^{-2}\right]\mathcal{H}\sum\limits_I(1+\omega_I)a^{1-3\omega_I}\boxed{\left(A_I\tilde {\bf
v}_I-\nabla\xi_I\right)}\nn\\
&+& \frac{2\kappa}{a^2}\left(k^2+\frac{4a^2}{3\lambda^2}\right)^{-1}\sum\limits_I(1+\omega_I)\left[a^{1-3\omega_I}\boxed{\left(A_I\tilde {\bf
v}_I-\nabla\xi_I\right)}\right]'\nn\\
&=&-\left(2k^2+\frac{4}{3a}\frac{d}{da}\left[\frac{a^4}{\lambda^2}\right]\right)\left(k^2+\frac{4a^2}{3\lambda^2}\right)^{-1}\mathcal{H}\hat{\bf B}_{\Sigma
I}\nn\\
&+& \frac{2\kappa}{a^2}\left(k^2+\frac{4a^2}{3\lambda^2}\right)^{-1}\sum\limits_I(1+\omega_I)a^{1-3\omega_I}\overline A_I\left[\hat{\bf
B}'+\mathcal{H}(1-3\omega_I)\hat{\bf B}\right]\, .\ea

Eq. \rf{3.17} for the ``J''-components in momentum space gives
\ba{4.9} \hat{\bf B}_{\Sigma J}=\frac{2\kappa}{a^2}\left(k^2+\frac{4a^2}{3\lambda^2}\right)^{-1}\sum\limits_J(\overline\varepsilon_J+\overline
p_J)a^4\boxed{\left(\tilde {\bf v}_J-\nabla \nu_J\right)}\, .
\ea
Further, we split the conservation equation \rf{a.23} for the ``J''-components (with the help of Eq. \rf{2.26}) into gradient and curl parts:
\be{4.10}
-[(\overline\varepsilon_J+\overline p_J)\nu_J]'-4\mathcal{H}(\overline\varepsilon_J+\overline p_J)\nu_J -\delta p_J-(\overline\varepsilon_J+\overline
p_J)\Phi=0
\ee
and
\be{4.11} [(\overline\varepsilon_J+\overline p_J){\bf B}]'-[(\overline\varepsilon_J+\overline p_J)(\tilde{\bf
v}_J-\nabla\nu_J)]'+4\mathcal{H}(\overline\varepsilon_J+\overline p_J){\bf B}-4\mathcal{H} (\overline\varepsilon_J+\overline p_J) (\tilde{\bf
v}_J-\nabla\nu_J)=0\, . \ee
Then, we obtain from Eqs. \rf{4.9} and \rf{4.11}
\ba{4.12} \hat{\bf B}_{\Sigma
J}'&=&-\left(2k^2+\frac{4}{3a}\frac{d}{da}\left[\frac{a^4}{\lambda^2}\right]\right)\left(k^2+\frac{4a^2}{3\lambda^2}\right)^{-1}\mathcal{H}\hat{\bf B}_{\Sigma
J}\nn\\
&+& \frac{2\kappa}{a^2}\left(k^2+\frac{4a^2}{3\lambda^2}\right)^{-1}a^4\sum\limits_J\left[(\overline\varepsilon_J+\overline p_J)\hat{\bf
B}'+(\overline\varepsilon_J+\overline p_J)'\hat{\bf B}+4\mathcal{H}(\overline\varepsilon_J+\overline p_J)\hat{\bf B}\right]\, .\ea
Summing up Eqs. \rf{4.3}, \rf{4.8} and \rf{4.12}, we get
\ba{4.13}
&&\left(k^2+\frac{4a^2}{3\lambda^2}\right)\hat{\bf B}'=-\left(2k^2+\frac{4}{3a}\frac{d}{da}\left[\frac{a^4}{\lambda^2}\right]\right) \mathcal{H}\hat{\bf
B}+2\kappa a^2\left(\frac{\overline\rho_M c^2}{a^3}\mathcal{H}\hat{\bf B}+\frac{\overline\rho_M
c^2}{a^3}\hat{\bf B}'\right)\nn\\
&+& 2\kappa a^2\sum\limits_I\frac{(1+\omega_I)\overline A_I}{a^{3+3\omega_I}}\left[\hat{\bf
B}'+\mathcal{H}(1-3\omega_I)\hat{\bf B}\right]\nn\\
&+& 2\kappa a^2\sum\limits_J\left[(\overline\varepsilon_J+\overline p_J)\hat{\bf B}'+(\overline\varepsilon_J+\overline p_J)'\hat{\bf
B}+4\mathcal{H}(\overline\varepsilon_J+\overline p_J)\hat{\bf B}\right]\, .
\ea
Taking into account the definition \rf{3.3} for $\lambda$, it is not difficult to show that this equation reduces to the following one:
\be{4.14} \hat{\bf B}'+2\mathcal{H}\hat{\bf B}=0\, , \ee
which is the Fourier transform of Eq. \rf{4.1}.

\subsection{Check of Equation \rf{2.10} for $\Phi$}

Now we will prove Eq. \rf{2.10}. To do it, we first note that from Eq. \rf{2.28} one can get
\ba{4.15}
&&\Phi''+3{\mathcal H}\Phi'+\left({2\mathcal H}'+\mathcal{H}^2\right)\Phi = \left({\mathcal H}'-\mathcal{H}^2\right)\Phi\nn\\
&-&\frac{\kappa c^2}{2}\left(\frac{\Xi}{a}\right)'-\frac{\kappa}{2}\sum\limits_I\left(1+\omega_I\right)\left(\frac{\xi_I}{a^{1+3\omega_I}}\right)'-\frac{\kappa
}{2}\sum\limits_J\left[a^2(\overline\varepsilon_J+\overline p_J)\nu_J\right]'\nn\\
&-&\frac{\kappa c^2}{a}\mathcal{H}\Xi-\kappa\mathcal{H}\sum\limits_I\frac{1+\omega_I}{a^{1+3\omega_I}}\xi_I-\kappa
a^2\mathcal{H}\sum\limits_J(\overline\varepsilon_J+\overline p_J)\nu_J\, .
\ea
Therefore, to prove Eq. \rf{2.10}, it is sufficient to show that $\frac{1}{2}\kappa a^2\left(\sum\limits_I\delta{p}_I+\sum\limits_J\delta{p}_J\right)$ is equal to
the right-hand side of Eq. \rf{4.15}. We can write this condition as follows:
\ba{4.16}
&&\frac{1}{2}\kappa a^2\left(\sum\limits_I\delta{p}_I+\sum\limits_J\delta{p}_J\right)+\left(\mathcal{H}^2-{\mathcal
H}'\right)\Phi=-\frac{\kappa c^2}{2a}\left(\Xi'+\mathcal{H}\Xi\right) \nn\\
&-&\frac{\kappa}{2}\sum\limits_I\frac{1+\omega_I}{a^{1+3\omega_I}}\xi'_I+\frac{\kappa}{2}\sum\limits_I\frac{(1+\omega_I)(1+3\omega_I)}{a^{1+3\omega_I}}
\mathcal{H}\xi_I-\kappa\mathcal{H}\sum\limits_I\frac{1+\omega_I}{a^{1+3\omega_I}}\xi_I\nn\\
&-&\frac{\kappa a^2}{2}\sum\limits_J\left[(\overline\varepsilon_J+\overline p_J)\nu_J\right]'- 2\kappa
a^2\mathcal{H}\sum\limits_J(\overline\varepsilon_J+\overline p_J)\nu_J\, . \ea
To determine $\xi'$ and $\left[(\overline\varepsilon_J+\overline p_J)\nu_J\right]'$, we can use Eqs. \rf{4.6} and \rf{4.10}, respectively. Additionally, it is not
difficult to show from Eqs. \rf{3.9} and \rf{4.2} that $\Xi'+\mathcal{H}\Xi=-\bar{\rho}_M\Phi$, where we dropped the quadratic term $O(\tilde{v}^2_n)$ and used
the condition $\mathbf{k}\hat{\mathbf{B}}=0$. Then, we get
\ba{4.17}
&&\frac{1}{2}\kappa a^2\sum\limits_I\delta{p}_I+\left(\mathcal{H}^2-{\mathcal H}'\right)\Phi=\frac{\kappa c^2}{2a}\overline\rho_M
\Phi+\frac{\kappa}{2}\sum\limits_I\frac{(1+\omega_I)(1+3\omega_I)}{a^{1+3\omega_I}}\overline{A}_I\Phi\nn\\
&+&\frac{\kappa a^2}{2}\sum\limits_J(\overline\varepsilon_J+\overline p_J)\Phi
+ \frac{\kappa}{2}\sum\limits_I\frac{\omega_I\delta
A_I}{a^{1+3\omega_I}}\, .
\ea
Now, if we use Eq. \rf{2.15} to find $\delta p_I=\omega_I\delta\varepsilon_I$ and Eq. \rf{2.4} to determine $\mathcal{H}^2-\mathcal{H}'$, we can easily
demonstrate that Eq. \rf{4.17} is just an identity. Therefore, Eq.~\rf{2.10} is satisfied.

\section{Conclusion}

In our paper we have studied a universe filled with dust-like matter in the form of discrete inhomogeneities (e.g., galaxies and their groups and clusters) which
represents the CDM-component and additionally with  two groups of other matter sources which can be responsible for dark energy. To cover a wide class of cases,
we have considered a very general model where the first group of sources consists of perfect fluids with a linear EoS $p_I=\omega_I\varepsilon_I\, \
(\omega_I=\mathrm{const})$. The second group of matter sources consists of perfect fluids with an arbitrary nonlinear EoS: $p_J=f_J(\varepsilon_J)$. The
background spacetime geometry is defined by the FLRW metric.

We have developed the first-order scalar and vector cosmological perturbation theory. Our approach works at all cosmological scales and incorporates both linear
and nonlinear effects with respect to energy density fluctuations. The only restriction is that we consider the weak gravitational field limit. We have
demonstrated that the scalar perturbation $\Phi$ (i.e. the gravitational potential) as well as the vector perturbation $\mathbf{B}$ can be split into individual
contributions from each matter source: $\Phi=\Phi_M+\sum_I\Phi_I+\sum_J\Phi_J$ and $\mathbf{B}=\mathbf{B}_M+\sum_I\mathbf{B}_I+\sum_J\mathbf{B}_J$. Each of these
contributions satisfies its own equation (see Eqs. \rf{3.5}-\rf{3.7} and \rf{3.15}-\rf{3.17}). The velocity independent parts of $\Phi_M, \Phi_I$ and $\Phi_J$ are
characterized by the finite time-dependent Yukawa interaction range $\lambda$, defined by the formula \rf{3.3} and being the same for each individual
contribution. We have also obtained the exact form of $\Phi_M$ and $\mathbf{B}_M$ related to the discrete matter sources. We have performed a thorough check of
the self-consistency of our approach.

It is important to note that the equations obtained in our paper form
the theoretical basis for subsequent numerical simulations for a very
wide class of cosmological
models. Since our approach is valid at arbitrary cosmological scales, we can use these equations for studying the mutual motion of galaxies and the Hubble flow
formation at relatively small scales (e.g., up to 20-30 Mpc), as well
as for the investigation of structure formation at very large cosmological distances
1000-3000 Mpc corresponding to the largest known cosmic structures \cite{supstr1,supstr2,supstr3}.  The formation of such enormously large structures is a
challenge of modern cosmology because they considerably exceed the previously reported cell of homogeneity dimension $\approx 370$ Mpc \cite{Yadav}.

\section*{Acknowledgements}

The work of M. Eingorn was partially supported by an Albert's
Researcher Reunion Grant of the University of Cologne.

\appendix
\section{Energy-momentum tensors and conservation equations in the first-order approximation}
\renewcommand{\theequation}{A.\arabic{equation}}

\setcounter{equation}{0}

In this appendix we obtain expressions for the components of the perturbed energy-momentum tensors for perfect fluids. To start with, we determine the
contravariant components of the metric \rf{2.5} in the first-order (with respect to $\Phi$ and $B_{\alpha}$) approximation. The nonzero components are
\be{a.1}
g^{00}\approx a^{-2}(1-2\Phi), \quad g^{\alpha\alpha} \approx -a^{-2}(1+2\Phi), \quad g^{0\alpha} \approx a^{-2}B_{\alpha},\quad \alpha =1,2,3\, .
\ee
The energy-momentum tensor of perfect fluids reads
\be{a.2} T^{ik} = \left(\varepsilon + p\right)u^iu^k - p g^{ik}\, , \quad i,k = 0,1,2,3\, , \ee
where the four-velocity components are $u^i \equiv dx^i/ds$. Taking into account that in the first-order approximation\footnote{In this approximation, we keep the
metric corrections $\Phi, B_{\alpha}$ and the peculiar velocity $\tilde v^{\alpha}$ in the linear order. We do not consider strong gravitational fields, as it is
usually assumed for the cosmological problem setting. Therefore, e.g., $|\Phi|\ll 1$. It results, in particular, in the inequality $|\tilde v^{\alpha}\Phi| \ll
|\tilde v^{\alpha}|$. It is also not difficult to realize that the terms $\tilde v^{\alpha}\tilde v^{\beta}, \tilde v^{\alpha}\Phi$ and $\tilde
v^{\alpha}B_{\beta}$ on the right-hand side of the perturbed Einstein equations result in quadratic correction terms for the metric coefficients, which are beyond
the accuracy of the first-order approximation. Hence, we neglect these terms.} $ds/d\eta = a\left[(1+2\Phi)+2B_{\alpha}\tilde
v^{\alpha}-(1-2\Phi)\delta_{\alpha\beta}\tilde v^{\alpha}\tilde v^{\beta}\right]^{1/2} \approx a (1+\Phi)$ (where the peculiar velocity $\tilde v^{\alpha}\equiv
dx^{\alpha}/d\eta$), we get
\be{a.3} u^0 = \frac{d\eta}{ds} \approx \frac{1}{a}(1-\Phi), \quad u^{\alpha} = \frac{dx^{\alpha}}{d\eta}\frac{d\eta}{ds}\approx \tilde
v^{\alpha}\frac{1}{a}(1-\Phi)\approx \frac{\tilde v^{\alpha}}{a}\, . \ee
Then, the nonzero components of the energy-momentum tensor are
\ba{a.4}
T^{00}&\approx& \varepsilon\frac{1}{a^2}(1-2\Phi)\, ,\\
\label{a.5}
T^{0\alpha}&\approx& (\varepsilon+p)\frac{1}{a^2}\tilde v^{\alpha}-\frac{1}{a^2}p B_{\alpha}\, ,\\
\label{a.6}
T^{\alpha\beta}&\approx& \frac{1}{a^2}p(1+2\Phi)\delta_{\alpha\beta}\, ,\\
\label{a.7}
T^{0}_{0}&\approx&\varepsilon\, ,\\
\label{a.8}
T^0_{\alpha}&\approx& (\varepsilon+p)B_{\alpha}-(\varepsilon+p)\tilde v^{\alpha}\, ,\\
\label{a.9}
T^{\alpha}_0&\approx& (\varepsilon+p)\tilde v^{\alpha}+p B_{\alpha}\, ,\\
\label{a.10}
T^{\alpha}_{\beta}&\approx&-p\delta_{\alpha\beta}\, .
\ea
The energy-momentum tensor for discrete gravitating sources of masses $m_n$ can be written in the form \cite{Landau}
\be{a.11}
T^{ik} = \sum_n \frac{m_n c^2}{\sqrt{-g}}\frac{dx^i_n}{d\eta}\frac{dx^k_n}{d\eta}\frac{1}{ds_n/d\eta}\delta({\bf r}-{\bf r}_n)\, .
\ee
We can introduce the rest mass density
\be{a.12}
\rho_M\equiv \sum_n m_n \delta({\bf r}-{\bf r}_n) \equiv \sum_n \rho_n\,
\ee
and write down the nonzero components:
\ba{a.13}
T^{00}&\approx& \frac{\rho_M c^2}{a^5}(1+\Phi)\, ,\\
\label{a.14}
T^{0\alpha}&\approx&\frac{1}{a^5}\sum_n\rho_n c^2\tilde v^{\alpha}_n\, ,\\
\label{a.15}
T^0_0&\approx& \frac{\rho_M c^2}{a^3}(1+3\Phi)\, ,\\
\label{a.16}
T^0_{\alpha}&\approx& \frac{\rho_M c^2}{a^3}B_{\alpha} - \frac{1}{a^3}\sum_n \rho_n c^2 \tilde v^{\alpha}_n\, ,\\
\label{a.17}
T^{\alpha}_0&\approx& \frac{1}{a^3} \sum_n \rho_n c^2 \tilde v^{\alpha}_n\, .
\ea
Since $T^0_0\approx \varepsilon_M$, from \rf{a.13} and \rf{a.15} we get $T^{00}\approx (\varepsilon_M/a^2)(1-2\Phi)$ in accordance with Eq. \rf{a.4}. Obviously,
the expressions \rf{a.13}-\rf{a.17} for the dust-like matter component can be rewritten in the form \rf{a.4}-\rf{a.10} when we put $p=0$ and suppose the
following splitting of the energy density: $\varepsilon = \sum_n \rho_n c^2 (1+3\Phi)/a^3\equiv \sum_n\varepsilon_n$; for example, 
$T^{0\alpha}\approx \sum_n\varepsilon_n\tilde v^{\alpha}_n/a^2\approx \sum_n\rho_n c^2\tilde v^{\alpha}_n/a^5$.

The covariant energy-momentum conservation equations read:
\be{a.18}
T^{k}_{i;k}=\frac{1}{\sqrt{-g}}\frac{\partial \left(T_{i}^{k}\sqrt{-g}\right)}{\partial x^{k}}-\frac{1}{2}\frac{\partial g_{kl}}{\partial x^{i}}T^{kl}=0\,
\, .
\ee
First, we consider the $i=0$ component of this equation. For the energy-momentum tensor components \rf{a.4}-\rf{a.10}, we can write it in the form
\ba{a.19}
&{}&\frac{1}{\sqrt{-g}}\frac{\partial \left(T_{0}^{0}\sqrt{-g}\right)}{\partial x^{0}}+\frac{1}{\sqrt{-g}}\frac{\partial
\left(T_{0}^{1}\sqrt{-g}\right)}{\partial x^{1}}+\frac{1}{\sqrt{-g}}\frac{\partial \left(T_{0}^{2}\sqrt{-g}\right)}{\partial
x^{2}}+\frac{1}{\sqrt{-g}}\frac{\partial \left(T_{0}^{3}\sqrt{-g}\right)}{\partial x^{3}}\nn\\
&-&\frac{1}{2}g'_{00}T^{00}-\frac{1}{2}g'_{11}T^{11}-\frac{1}{2}g'_{22}T^{22}-\frac{1}{2}g'_{33}T^{33}=0\, ,
\ea
where we took into account that $T^{0\alpha}\, (\alpha=1,2,3)$ are already linear expressions with respect to $\tilde v^{\alpha}$ and, being multiplied by
$g_{0\alpha}'$, they result in the second order. Therefore, we dropped such second-order terms. Substituting \rf{a.4}-\rf{a.10}, we get in the first-order
approximation:
\be{a.20}
\varepsilon'+3\mathcal{H}(\varepsilon+p)-3(\varepsilon+p)\Phi'+\nabla[(\varepsilon+p)\tilde{\bf v}]+\nabla[p{\bf B}]=0\,
.
\ee
Next, we consider the $i=\alpha$ components. We suppose for definiteness $\alpha=1$. Then, in the first-order approximation we obtain
\ba{a.21}
&{}&\frac{1}{\sqrt{-g}}\frac{\partial \left(T_{1}^{0}\sqrt{-g}\right)}{\partial x^{0}}+\frac{1}{\sqrt{-g}}\frac{\partial
\left(T_{1}^{1}\sqrt{-g}\right)}{\partial x^{1}}\nn\\
&-&\frac{1}{2}\frac{\partial g_{00}}{\partial x^{1}}T^{00}-\frac{1}{2}\frac{\partial g_{11}}{\partial x^{1}}T^{11}-\frac{1}{2}\frac{\partial g_{22}}{\partial
x^{1}}T^{22}-\frac{1}{2}\frac{\partial g_{33}}{\partial x^{1}}T^{33}=0\, ,
\ea
which for \rf{a.4}-\rf{a.10} reads
\ba{a.22} &{}& [(\varepsilon+p)B_1]'-[(\varepsilon+p)\tilde v^1]'+4\mathcal{H}(\varepsilon+p)B_1-4\mathcal{H}(\varepsilon+p)\tilde v^1 \nn\\
&-&\frac{\partial p}{\partial x^1}-(\varepsilon+p)\frac{\partial\Phi}{\partial x^1}=0\, . \ea
Consequently,
\ba{a.23}
[(\varepsilon+p){\bf B}]'-[(\varepsilon+p)\tilde{\bf v}]'+4\mathcal{H}(\varepsilon+p){\bf B}-4\mathcal{H}(\varepsilon+p) \tilde{\bf v} -\nabla
p-(\varepsilon+p)\nabla\Phi=0\, .
\ea

\section{Check of Equation \rf{2.28} for $\Phi$}
\renewcommand{\theequation}{B.\arabic{equation}}

\setcounter{equation}{0}

In this appendix, we demonstrate that Eq. \rf{2.28} is fulfilled for the scalar perturbation $\Phi$ with the components satisfying Eqs. \rf{3.5}-\rf{3.7}. This is
easy to show in momentum space. First, we obtain expressions for the time derivatives of the components $\hat \Phi_M, \hat \Phi_{\Sigma I}$ and
$\hat\Phi_{\Sigma J}$. From Eq. \rf{3.10} for $\hat\Phi_M$ we get
\ba{b.1} \hat\Phi'_M &=&
-\mathcal{H}\left(k^2+\frac{a^2}{\lambda^2}\right)^{-1}\left(a\frac{d}{da}\left[\frac{a^2}{\lambda^2}\right]+k^2+\frac{a^2}{\lambda^2}\right)
\hat\Phi_M-\frac{\kappa c^2}{2a}\left(k^2+\frac{a^2}{\lambda^2}\right)^{-1}\nn\\
&\times&\left[\sum\limits_nm_n\exp(-i{\bf k}{\bf r}_n)\left(-i({\bf k}\tilde{\bf v}_n)+3i\left[\frac{{\mathcal H}}{a}\right]'\frac{a({\bf k}\tilde{\bf
v}_n)}{k^2}\right)+3\mathcal{H}\overline\rho_M\hat\Phi\right]\, ,
\ea
where we used Eq. \rf{4.2} (keeping in mind that $\mathbf{k}\hat{\mathbf{B}}=0$) and dropped the term $O(\tilde{v}_n^2)$. From Eqs. \rf{3.6} and \rf{3.7} we get
expressions for the ``I''- and ``J''-components in momentum space:
\ba{b.2} \hat\Phi_{\Sigma I}=-\left(k^2+\frac{a^2}{\lambda^2}\right)^{-1}\left[\frac{\kappa a^2}{2}\sum\limits_I\frac{\hat{\delta
A}_I}{a^{3(1+\omega_I)}}-\frac{3\mathcal{H}\kappa}{2}\sum\limits_I\frac{1+\omega_I}{a^{1+3\omega_I}}\hat\xi_I\right]\, \ea
and
\ba{b.3}
\hat\Phi_{\Sigma J}=-\left(k^2+\frac{a^2}{\lambda^2}\right)^{-1}\left[\frac{\kappa
a^2}{2}\sum\limits_J\overline\varepsilon_J\hat\delta_J-\frac{3\mathcal{H}\kappa a^2}{2}\sum\limits_J(\overline\varepsilon_J+\overline p_J)\hat\nu_J\right]\, .
\ea
Then, the time derivative of Eq. \rf{b.2} gives
\ba{b.4}
\hat\Phi'_{\Sigma I}
&=&-\mathcal{H}\left(k^2+\frac{a^2}{\lambda^2}\right)^{-1}\left(a\frac{d}{da}\left[\frac{a^2}{\lambda^2}\right]+k^2+\frac{a^2}{\lambda^2}\right)
\hat\Phi_{\Sigma I}-
\left(k^2+\frac{a^2}{\lambda^2}\right)^{-1}\nn\\
&\times&\sum\limits_I\left[\frac{\kappa(1+\omega_I)}{2}\frac{k^2\hat\xi_I}{a^{1+3\omega_I}}-
\frac{3\kappa(1+\omega_I)}{2}\left[\frac{\mathcal{H}}{a^{1+3\omega_I}}\right]'\hat\xi_I\right.\nn\\
&+&\left. \frac{3\mathcal{H}\kappa}{2}\left(-3\omega_I\mathcal{H}\frac{(1+\omega_I) }{a^{1+3\omega_I}}\hat\xi_I +(1+3\omega_I)\frac{(1+\omega_I)\overline
A_I}{a^{1+3\omega_I}}\hat\Phi\right)\right]\, ,
\ea
where we used Eqs. \rf{2.16} and \rf{2.25} (e.g., $\delta A'_I=-(1+\omega_I)\nabla(A_I\tilde{\mathbf{v}}_I)=-(1+\omega_I)\triangle\xi_I \rightarrow \hat{\delta
A}'_I=k^2(1+\omega_I)\hat{\xi}_I$) and the expression for $\hat\xi'_I$ was obtained from Eq. \rf{4.6}.

Similarly
\ba{b.5}
\hat\Phi'_{\Sigma J}
&=&-\mathcal{H}\left(k^2+\frac{a^2}{\lambda^2}\right)^{-1}\left(a\frac{d}{da}\left[\frac{a^2}{\lambda^2}\right]+k^2+\frac{a^2}{\lambda^2}\right)\hat\Phi_{\Sigma
J}-
\left(k^2+\frac{a^2}{\lambda^2}\right)^{-1}\nn\\
&\times&\sum\limits_J\left[-\frac{3\mathcal{H}'\kappa
a^2}{2}(\overline\varepsilon_J+\overline p_J)\hat\nu_J-3\mathcal{H}\frac{\kappa a^2}{2}\left(\frac{d\overline
p_J}{d\overline\varepsilon_J}\overline\varepsilon_J\right)\hat\delta_J\right.\nn\\
&+&\left. \frac{\kappa a^2}{2}(\overline\varepsilon_J+\overline p_J)k^2\hat\nu_J+\frac{3\mathcal{H}\kappa
a^2}{2}\left(\mathcal{H}(\overline\varepsilon_J+\overline p_J)\hat\nu_J +\hat{\delta p}_J+(\overline\varepsilon_J+\overline p_J)\hat\Phi\right) \right]\, ,\ea
where we took into account Eqs. \rf{2.19}, \rf{2.22}, \rf{2.26} and \rf{4.10}.

Summing up Eqs. \rf{b.1}, \rf{b.4} and \rf{b.5}, we have
\ba{b.6}
&&\left(k^2+\frac{a^2}{\lambda^2}\right)(\hat\Phi' + \mathcal{H}\hat\Phi)=-\frac{\kappa c^2}{2a}\left[\sum\limits_nm_n\exp(-i{\bf k}{\bf
r}_n)\left(-i({\bf k}\tilde{\bf v}_n)+3i\left[\frac{{\mathcal H}}{a}\right]'\frac{a({\bf
k}\tilde{\bf v}_n)}{k^2}\right)\right]\nn\\
&-&\sum\limits_I\left[\frac{\kappa(1+\omega_I)}{2}\frac{k^2\hat\xi_I}{a^{1+3\omega_I}}-
\frac{3\kappa(1+\omega_I)}{2}\left[\frac{\mathcal{H}}{a^{1+3\omega_I}}\right]'\hat\xi_I+
\frac{3\mathcal{H}\kappa}{2}\left(-3\omega_I\mathcal{H}\frac{(1+\omega_I) }{a^{1+3\omega_I}}\hat\xi_I\right)\right]\nn\\
&-&\sum\limits_J\left[\mathcal{H}\frac{3\mathcal{H}\kappa a^2}{2}(\overline\varepsilon_J+\overline p_J)\hat\nu_J-\frac{3\mathcal{H}'\kappa
a^2}{2}(\overline\varepsilon_J+\overline p_J)\hat\nu_J+\frac{\kappa a^2}{2}(\overline\varepsilon_J+\overline p_J)k^2\hat\nu_J \right]\, ,
\ea
where we used Eqs. \rf{2.21} and \rf{3.3}. Keeping in mind that $k^2\hat\Xi=-i\sum\limits_nm_n({\bf
k}\tilde{\bf v}_n)\exp(-i{\bf k}{\bf r}_n)$ (see Eq. \rf{3.9}) and $(\mathcal{H}/a)'=(\mathcal{H}'-\mathcal{H}^2)/a=-a/(3\lambda^2)$ (see Eq. \rf{2.4}), we get
\ba{b.7}
&&\left(k^2+\frac{a^2}{\lambda^2}\right)(\hat\Phi' + \mathcal{H}\hat\Phi)=-\frac{\kappa c^2}{2a}\left(k^2+\frac{a^2}{\lambda^2}\right)\hat\Xi\nn\\
&-&\sum\limits_I\left[\frac{\kappa(1+\omega_I)}{2}\frac{\hat\xi_I}{a^{1+3\omega_I}}\left(k^2+\frac{a^2}{\lambda^2}\right)\right]-
\sum\limits_J\left[\frac{\kappa
a^2}{2}(\overline\varepsilon_J+\overline p_J)\left(k^2+\frac{a^2}{\lambda^2}\right)\hat\nu_J \right]\, .
\ea
Obviously, this equation results in Eq. \rf{2.28}.

\end{document}